# Application of Lowner-John Ellipsoid in the Steganography of Lattice Vectors

Hossein Mohades[1*], Mohamad Kadkhoda[2] and Mohamad Mahdi Mohades[3]

1. Mathematics and Computer ScienceDepartment ,Amirkabir University, Vali-e-asr,Tehran, Tehran, Iran.
2. Mathematical and Informatics Research Group, Tarbiat Modares University, north Kargar, Tehran, Tehran, Iran.
3. Electrical Engineering Department, Iran University of Science and Technology, Narmak, Tehran, Tehran, Iran.

*Corresponding author: hoseinmohades@yahoo.com

**Abstract**
In this paper, first, we utilize the Lowner-John ellipsoid of a convex set to hide the lattice data information. We also describe the algorithm of information recovery in polynomial time by employing the Todd-Khachyian algorithm. The importance of lattice data is generally due to their applications in the homomorphic encryption schemes. For this reason we also outline the general scheme of a homomorphic encryption provided by Gentry.

**Keywords:** *Lattice, Lowner-John ellipsoid, Steganography, Homomorphic cryptography, Todd-Khachyian algorithm.*

## 1. Introduction

The classical model for invisible communication was firstly presented by Simmons [8]. Alice and Elena are imprisoned. They want to make an escape plan. Turnkey fully supervises them, and any suspicious message would corrupt their relationship. Suspicious statements include statements code, which might be inaudible. So they ought to situate their important information inside a normal text patterns to not arouse suspicion. This is called steganography [1,7,8]. For example, Elena can send Alice a post-modern painting, data cover space, and the turnkey cannot understand whether the painter follows any special purpose through the geometrical forms.

The first applicable version of the Fully Homomorphic Encryption was introduced by Gentry in his PhD thesis and then improved by Smart, Vercauteren, Gentry and Halevi in [3,4,5,13]. The public key of these schemes corresponds to a basis (skewed basis) for a lattice, while the private key corresponds to another more orthogonal basis (good basis) of the same lattice.

A public-key steganography allows two people, who have never exchanged a secret, to send hidden messages over a public channel so that an adversary cannot even detect that these hidden messages are being sent [12].

Our goal in this paper is to introduce a simple way to send secret information, such as the vectors of a lattice vector, using a new steganography method. This method is applicable for binary data, media as well as vectors.

For this purpose, John's uniqueness Ellipsoid theorem and Edges Based data Embedding steganography method are exploited. We hide the vectors information via the maximal length vector of the ellipsoid associated to a convex polytope. We also describe the algorithm of information recovery in polynomial time with the help of the Todd-Khachyian algorithm.

This method of steganography is important to improve the security of the encryptions based on the lattices like the Gentry's fully homomorphic encryption. For the sake of completeness and educational purposes we provide a description of a toy model for the fully homomorphic cryptography scheme, based on notes from the Gentry and Halevi speeches and papers [3,4,5,6,10]

## 2. Description of the Gentry's scheme for fully homomorphic encryption.

The basic of fully homomorphic encryption is somewhat homomorphic encryption. Somewhat homomorphic encryption is a type of encryption which is homomorphic for a special class of functions. This type of encryption is suitable for computing the problems presented by polynomials of small degree.

The first semantic secure fully homomorphic encryption was presented by Craig Gentry [3,4].

He extend the domain of a somewhat homomorphic encryption which was introduced by Cracker to a broader domain using the method of bootstrapping. We provide a description of the Cracker's method here. Cracker applied a set of polynomials $f : Z_q^n \to Z_q$ which are zero at the point $(s_1,\ldots,s_n) \in Z_q^n$ as an encryption of the element $0 \in Z_q^n$. This set is denoted by $Z(0)$. In fact he used the set $Z(0)$ as the public key of his encryption. The special key is the point $(s_1,\ldots,s_n) \in Z_q^n$. To encrypt an element $m$ one chooses a random element $g \in Z(0)$, the ciphertext is $c(x_1,\ldots,x_n) = m + g(x_1,\ldots,x_n)$ to decrypt the ciphertext it is enough to use the special key $(s_1,\ldots,s_n)$ and compute the evaluation of $c(x_1,\ldots,x_n)$ at this point.

$$Dec_{sk}(c(x_1,\ldots,x_n)) = \\ Eval(sk, c(x_1,\ldots,x_n)) = c(s_1,\ldots,s_n) \quad (1)$$

It is easy to see that the Cracker's encryption is a fully homomorphic encryption. But unfortunately it is not semantically secure. The problem of semantic security of Cracker's encryption is equivalent to the problem of Ideal's membership. So, it is easy to brick the Cracker's encryption. Gentry added some noises to the Cracker's encryption to overcome this problem. Note that we can use binary numeral system, therefore it is enough to find a fully homomorphic encryption of 0 and 1.

Let us represent $Z_p$ with the set $\{0,1,\ldots,p-1\} \subset \mathbb{Z}$. Gentry used the set of all polynomials

$$SE(0) = \{f : f_i(s_1,\ldots,s_n) = 2e_i \ s.t. |e_i| << p\}$$

as the new public Key. These polynomials are nominated as Small and even polynomials or smeven. For encryption one must choose a random element $g \in SE(0)$. The ciphertext associated to $m \in \{0,1\}$ is the polynomial

$$c(x_1,\ldots,x_n) = m + g(x_1,\ldots,x_n) \quad (2)$$

For decryption it is enough to evaluate $c(x_1,\ldots,x_n)$ at the secret key $(s_1,\ldots,s_n)$ and then compute it in mod 2. $c(s_1,\ldots,s_n)$ is called the noise of the ciphertext.

Now we overcame the security problem of the Cracker's encryption. But, still there is a problem.

The noise $2e_i$ grows under summations and multiplications. After some multiplication and summation it is possible that $m_1 + m_2 + \ldots + m_k + 2(e_1 + e_2 + \ldots + e_k)$ be a multiple of $p$ where $m_1 + m_2 + \ldots + m_k$ is an odd number. So, we have an error in our computation. So the algorithm is true only for special polynomial of small degree. In fact we have a somewhat homomorphic encryption.

### 3. Somewhat homomorphic encryption using polynomials of degree 1.

In this section we apply the Cracker's algorithm for polynomials of degree 1. This helps us to have a less sophisticated but secure codes which are suitable for bootstrapping techniques. Let $q$ be a natural number such that $\gcd(q,2) = 1$. We use an arbitrary element $(s_1,\ldots,s_n)$ of $Z_q^n$ as the special key (uniform distribution). Let,

$$c(x_1,\ldots,x_n) = m + g(x_1,\ldots,x_n) \quad (3)$$

be a set of functions of degree 1 such that

$$f_i(s_1,\ldots,s_n) = 2e_i \quad |e_i| \ll q . \quad (5)$$

Let $g = a_0 + a_1 x_1 + \ldots + a_n x_n$ be an arbitrary element of $S1(0)$. As in the Cracker's algorithm case let the ciphertext associated with $m$ be the polynomial $c(x) = m + g(x)$. The semantic security is a result of the Hash Lemma. Summation over polynomials $c_1 = m_1 + g_1(x), c_2 = m_2 + g_2(x)$ preserves the degree. But multiplication of polynomials $c_1 = m_1 + g_1(x), c_2 = m_2 + g_2(x)$ leads to a polynomial of degree 2.

$$c_1.c_2 = m_1 m_2 + m_1 g_2 + m_2 g_1 + g_1 g_2 \\ = \sum_{\{i,j=0\}}^{n} c_{1i} c_{2j} x_i x_j \quad (6)$$

where $x_0 = 1$.

Note that the cloud servers are able to handle huge data. But, decryption of data's has its own limits. Growing of the degrees under the multiplication is a big problem. Now we present a solution for this problem.

$c_1.c_2$ is a polynomial of degree 2. We use a re-encryption of $c_1.c_2$ via another secret key in such a way that the noise part of multiplication looks like a linear function. First using reparametrizations $S_{ij} = s_i s_j$ and $X_{ij} = x_i x_j$ of secret key and variables we achieve a degree one polynomial of $O(n^2)$ variables.

$$C(X) = c_1(x)c_2(x) = \sum_{ij} c_i c_j X_{ij} = \sum_{ij} C_{ij} X_{ij}. \quad (7)$$

Now we show that there is a way to transform a long linear ciphertext $C(X)$ with $N > n$ variables to a shorter one. We ask the cloud server to send us the function $d(x) = \sum_{ij} c_i c_j h_{ij}(x)$ where $\{h_{ij}(x)\}$ are linear functions of $n$ variables which are added to the public Key. $h_{ij}(x)$ is chosen in such a way that $h_{ij}(t) = S_{ij}$, for a second secret Key $= (t_1, \ldots, t_n)$. The answer of the cloud server is now a linear function of $n$ variables (instead of a linear function of $O(n^2)$ variables). So, we must evaluate a linear function of $n$ variables to decrypt the multiplication $c_1 c_2$. This is a toy model of the bootstrapping method.

There is only a small problem. Let us have a glance at the next formula

$$d(t) = \sum_{ij} C_{ij} \cdot (S_{ij} + smeven_{ij})$$
$$= C(S) + \sum_{ij} C_{ij} \cdot smeven_{ij}. \quad (8)$$

It is possible that $\sum_{ij} C_{ij} \cdot smeven_{ij}$ is not a smeven. Gentry applied the method of bit decomposition for $C(X)$ to replace $C_{ij}$ with $C_{ij}'$ such that

$$C(S) + \sum_{ij} C_{ij} \cdot smeven_{ij} ((mod\, p) mod\, 2)$$
$$((mod\, p) mod\, 2) \quad (9)$$
$$= C(S) + \sum_{ij} C_{ij}' \cdot smeven_{ij}$$

and $\sum_{ij} C_{ij}' \cdot smeven_{ij}$ be an smeven. So, using this action (which is called the $SwitchKey(s,t)$ action) we can control the increasing of noises. And therefore we have an effective fully homomorphic encryption. In a similar way phenomenon like before leads to a lattice based fully homomorphic encryption. Now we describe an LPR encryption. An LPR encryption consists of a parameter $q$, a ring $R = Z_q[y]/y^n + 1$, a secret key $s \in R$, and a set of functions $L(0) = \{f_i(x)|f_i(s) = 2e_i, |e_i| \ll q\}$. For encryption of an element $m \in Z_2[y]/y^n + 1$ we add an arbitrary element $g \in L(0)$ to $m$. A phenomenon like before leads to a lattice based fully homomorphic encryption.

**4. Lattice data steganography**

The proposed method of data steganography is based on Edges Based data Embedding (EBE) method. In order to hide an ordered set of vectors $v_1, \ldots, v_n \in R^n$ we utilize distinct vector steganography $(i, v_i) \in R^n$. Now we provide some definitions in functional analysis which are used in this section.

**Definition 4-1**: A subset of a vector space is convex when
$$x, y \in V \rightarrow tx + (1-t)y \in V \quad 0 \le t \le 1. \quad (10)$$

**Definition 4-2**: In a metric space, a set is compact provided that any Cauchy sequence of this set owns a convergent subsequence [9].

**Definition 4-3**: Suppose that $K \subseteq R^n$ is an n-dimensional compact convex set. Among all ellipsoids enclosing $K \subseteq R^n$, there is a unique ellipsoid of minimal volume which is called the John-Lowner ellipsoid of the convex set [2].

We aim to hide the important vectors of the lattice in a convex set. To this end, we set these vectors as the principal vectors of the John-Lowner ellipsoid of the convex object.

**Definition 4-4**: An embedding process is a function defined as $E : C \times M \rightarrow C$, where, $C$ is a set of covers and $M$ is a set of utilized texts. The recovery function $D$, $D : C \rightarrow M$, recovers original text from the cover space. In this case both transmitter and receiver must have access to the embedding and recovery functions. When $D(E(c,m)) = m$, then $(C, M, D, E)$ is called a pure steganography system.

In order to create our steganography scheme we act as follows. Let $M_C$ be the set of all compact and convex subsets in $R^n$, and C be the set of all ellipsoids included in $R^n$. Considering the fact that any ellipsoid is its own John's ellipsoid, all ellipsoids can be conceived as the John's ellipsoid of some compact convex sets. The set of compact convex spaces, whose John's ellipsoid is E, is presented by $C_E$. In order to accomplish steganography of an ordered set of vectors $v_1,\ldots,v_s \in R^{n-2}$, we use distinct steganography of vectors $(i, j_i, v_i) \in R^n$, where, $j_i$ indicates the coordinates system part in which the vector $v_i$ is located. To perform steganography, firstly, we hide a vector in the ellipsoid E and then, we hide this ellipsoid by selecting one member of $C_E$.

**Theorem 4-5**: For any vector there exists a set of ellipsoids centered at origin where the vector $v$ and its reflection $-v$ are the only vectors with the longest distance from the origin.

Proof: The set of $n$-dimensional ellipsoids is invariant under the action of orthogonal group $SO(n+1)$. There is a matrix $A$ in $SO(n+1)$ which sends a given vector $w$ of length $\|v\|$ to the vector $v$. Also, $A(-w) = -v$. Now, using a matrix $B \in SO(n+1)$ which sends $(\|v\|, 0, \ldots, 0)$ to $v$ on the ellipsoid $(\|v\|^2)x_1^2 + \alpha(\sum_{(i>1)} x_i^2) = 1$, where $\alpha > \frac{1}{\|v\|}$. The appropriate ellipsoid is prepared. □

The set of ellipsoids centered at origin is denoted by $B$ and the ellipsoids centered at origin for which the only maximum length vectors are $v, -v$ is denoted by $B_v$. To distinguish the selection of $x, -x$ we use the integer $j_i \in \{1, 2, \ldots, 2^{n-2}\}$ which illustrates the direction of vector. Now, we suppose that the members of vector space $V$ are the text elements which must be hidden, i.e. $M = V$. Moreover assume that the covering set is the set of subsets of $M_C$, i.e. $C = P(M_C)$. For the function $f : V \to B$, which acts as the element $v$ in $B_v$, and the function $g : B \to M_C$, which plays the role of ellipsoid $E$ in the set $C_E$, we define the following steganography embedding function:

$$P(M_C) \times V \to P(M_C) : (A, T)$$
$$\xrightarrow{L} \left(\frac{T}{s}\right) \cup \{gof(A)\} \quad . \quad (11)$$

Where $s$ is selected so that $gof(A)$ is greater than all ellipsoids. By defining the function $e$ as a function which sends the compact convex set $T$ to the largest ellipsoid included in $T$, we have the following theorem:

**Theorem 4-6**: $(L, e, P(M_C), V)$ is a pure steganography system.

Proof: By the axiom of choice and the theorem 4-5 there is a 1-1 map from $V$ to $P(M_c)$. Therefore, $card(V) \geq card(P(M_c))$. Also, looking at the unique vector of maximal norm in an element of $P(M_c)$ we have $D(E(c,m)) = m$. As a result, $(L, e, P(M_C), V)$ is a pure steganography system. □

Note that in order to provide a space to present effective algorithms; we may replace the steganography set constructed by all convex spaces with the set generated by convex polyhedrons. In fact, any convex polyhedron can be expressed by finite set of numbers. Using these numbers, the polyhedron can be constructed. Also note that instead of using the convex set $Y$ it is possible to utilize the set whose convex hull is $Y$.

## 5. Extraction Algorithm

In this section we provide an approximate extraction algorithm for over steganography plan, when the steganography set is a set of convex polyhedrons. This extraction algorithm is based on the Khachiyan-Todd ellipsoid finding algorithm [11].

Let $A$ denotes both $A = \{a^1, \ldots, a^m\}$ and the matrix constructed by them. The first part of the algorithm provides a subset $X_0 \subseteq A$ which has better behavior to find the Lowner-John ellipsoid.

---

*Let* $\mathfrak{A} = \{A_1, \ldots, A_s\}$ *be the set of all convex*
*polygons in the steganography plan*
*for* $h = 1 : s$
$A = A_h$
$A = \{a^1, \ldots, a^m\} \subset \mathbb{R}^n$
*if* $n \leq m \leq 2n$, *then* $X_0 \leftarrow A.v.$ Re*turn*
$\psi \leftarrow \{0\}, X_0 \leftarrow \varnothing, k \leftarrow 0.$
*while* $\mathbb{R}^n \setminus \psi \neq \varnothing$, *do*
*loop*
$k \leftarrow k+1$; *pick an arbitrary direction* $b^k \in \mathbb{R}^n$ *in*
*the orthogonal complement of* $\psi$
$\alpha \leftarrow \arg\max_{i=1,\ldots,m} (b^k)^T a^i \; X_0 \leftarrow X_0 \cup \{a^\alpha\}$
$\beta \leftarrow \arg\max_{i=1,\ldots,m} (b^k)^T a^i \; X_0 \leftarrow X_0 \cup \{a^\beta\}$
$\psi \leftarrow span\{\psi, a^\beta - a^\alpha\}$
*endloop*

*Let $p^0 \in \mathbb{R}^m$ be such that $p_i^0 = 1/|X_0|$, for $a^i \in X_0$, $p_i^0 = 0$ otherwise*

*$k \leftarrow 0, d \leftarrow n+1, q^i \leftarrow ((a^i)^T, 1)^T, i = 1,...,m.$*

*$\varepsilon_0 \leftarrow \{y \in \mathbb{R}^n : y^T \Lambda(p^0)^{-1} y \leq 1\}$*

*while does not satisfies the $\eta$-approximate optimaly conditions, do*

*loop*

$j_+ \leftarrow \arg\max \{(q^i)^T \Lambda(p^k)^{-1} q^i : i = 1,...,m\},$

$\kappa_+ \leftarrow (q^{j_+})^T \Lambda(p^k)^{-1} q^{j_+};$

$j_- \leftarrow \arg\max \{(q^i)^T \Lambda(p^k)^{-1} q^i : i = 1,...,m, p_i^k > 0\},$

$\kappa_- \leftarrow (q^{j_-})^T \Lambda(p^k)^{-1} q^{j_-};$

$\varepsilon_+ \leftarrow (\kappa_+/d) - 1, \varepsilon_- \leftarrow -(\kappa_-/d) + 1$

$\varepsilon_k \leftarrow \{\varepsilon_+, \varepsilon_-\}$

*if $\varepsilon_+ = \varepsilon_-$ then*

$\beta_k \leftarrow \frac{\kappa_+ - d}{n(\kappa_+ - 1)};$

$p^{k+1} \leftarrow (1-\beta_k) p^k + \beta_k e^{j_+}, k \leftarrow k+1;$

*else* $\beta_k \leftarrow \{\frac{-\kappa_- + d}{n(\kappa_- - 1)}, \frac{p_{j_-}^k}{1 - p_{j_-}^k}\}$

$p^{k+1} \leftarrow (1-\beta_k) p^k + \beta_k e^{j_-}, k \leftarrow k+1;$

$p^{k+1} \leftarrow (1-\beta_k) p^k + \beta_k e^{j_+}, k \leftarrow k+1;$

*else*

$\beta_k \leftarrow \{\frac{-\kappa_- + d}{n(\kappa_- - 1)}, \frac{p_{j_-}^k}{1 - p_{j_-}^k}\}$

$p^{k+1} \leftarrow (1-\beta_k) p^k + \beta_k e^{j_-}, k \leftarrow k+1;$

$\beta_k \leftarrow \frac{\kappa_+ - d}{n(\kappa_+ - 1)};$

$p^{k+1} \leftarrow (1-\beta_k) p^k + \beta_k e^{j_+}, k \leftarrow k+1;$

*else*

$\beta_k \leftarrow \{\frac{-\kappa_- + d}{n(\kappa_- - 1)}, \frac{p_{j_-}^k}{1 - p_{j_-}^k}\}$

$p^{k+1} \leftarrow (1-\beta_k) p^k + \beta_k e^{j_-}, k \leftarrow k+1;$

*and if*

$\varepsilon_k \leftarrow \{y \in \mathbb{R}^n : y^T \Lambda(p^k)^{-1} y \leq 1\}$

$X_k \leftarrow \{a^i \in A : p_i^k > 0\}\}.$

*endloop*

*Output $\varepsilon_k, X_k$*

$C(h, :) = \varepsilon_k$

$h = h + 1$

*end*

*output C*

---

The limit of the sets $X_k$ is $X_0$ and the limit of the ellipsoids $\varepsilon_k$ is the wanted Lowner-John ellipsoid. The matrix C provides the information of the john ellipsoids.

Since $\log vol \epsilon_{k+1}$ is equal to half of $\log \det \Lambda p^{k+1}$, the value of $\beta_k$ obtained with the help of the expression

$$\beta_k = \arg\max_{\beta \in [0,1]} (\log \det (1-\beta) \Lambda p^{k+1} + \beta e^{j_+}).$$

**Note 1**: According to the Khachiyan-Todd method up to a solution of a quadratic equation, the complexity of the extraction algorithm is $O(kmn^4(n\log(n) + n/\epsilon))$ for a $(1+\eta)$-approximation of the ellipsoid finding problem, where $k$ is the number of polytopes which are used in the steganography plan and $(1+\eta) = (1+\varepsilon)^{\frac{n}{2}}$. In fact, the first part of the given algorithm provides $(1+\eta)$-approximate quadratic form for a given convex set. Todd-Yildrim-Khachian says that the complexity of computation of each action of this approximation is $O(n\log(n) + n/\varepsilon)$. Also each action takes $O(mn)$ operations where $mn$ is the size of the matrix constructed by $m$ vectors in $\mathbb{R}^n$. Also we must apply the algorithm for all $k$ polytopes of the steganography plan. According to Lagrange multiplier method, finding the diameter of the ellipsoid provided by this algorithm is the same as finding the solution of a quadratic equation and a set of $n$ linear equations. By Gauss elimination algorithm the complexity of finding solutions for $n$ linear equation is $O(n^3)$. As a result, up to solving a quadratic equation the complexity of the given algorithm is $O(kmn^4(n\log(n) + n/\epsilon))$.

**Note 2**: There are many other spatial steganography methods such as spread spectrum steganography, LSB steganography, pixel value differencing steganography, code based steganography, mapping based steganography and palette based steganography; see [14, 15, 17, 18]. We compare our result with two of these methods, i.e. spread spectrum steganography and LSB steganography. Spread spectrum spectography considers the problem of blind extracting data embedded over a wide band in a domain of a digital medium. One of the most popular algorithms of this kind is M-IGLS. When the complexity of each iteration of this algorithm is $O(2K^3 + 2LMK + K^2(3L+M) + L^2K)$, where $L$ is the host image parameter and $K$ and $M$ respectively denote the number of distinct message and length of each message [15]. To compare, we set $K = m$ and $M = n$. So in our notation the complexity of each iteration is $O(2m^3 + 2Lnm + m^2(3L+n) + L^2m)$.

This is a polynomial of degree 3 in $m$ and degree 1 in $n$. So the complexity of our algorithm is higher than the complexity of this algorithm where the length of the hidden data is large and therefore our method is more secure than this algorithm for such a data. On the other hand this algorithm has better security where the number of hidden data is large.

Now, we compare our method with the LSB method.

LSB method, the Least Significant Bit technique, is the simplest and most popular method of image steganography. In many cases the key-recovery complexity of LSB method is asymptotically a linear function [16]. As a result, the security of our method is very high, however finding the keys are very hard even for friends. In fact, our method is totally different from the previous methods of steganography and it is natural to use this kind of steganography for a few numbers of massages of large length like the special keys of a fully homomorphic encryption or credit cards accounts.

## 6. Conclusion

In this paper we applied Lowner-John's ellipsoid to determine a steganography method to hide the messages containing the vectors of the lattices which is important to improve the security of the encryptions based on the lattices. Also we provide a review of the Gentry's idea for a fully homomorphic encryption. As future work we intend to develop algorithms to find the Lowner-John's ellipsoid for "special" convex sets which may result in implementation of the proposed method.

**References**

[1] Channalli, S. & Jadhav, A. (2009). Steganography an art of hiding, data International Journal on Computer Science and Engineering Vol.1 (3), 137-141.

[2] John, F. (1948). Extremum problems with inequalities as subsidiary conditions, Studies and Essays Presented to R. Courant on his 60'th Birthday, Interscience Publishers, Inc., New York.

[3] Gentry, C. (2009). A Fully Homomorphic Encryption Scheme, Ph.D. thesis

[4] Gentry, C. (2012). Winter school on cryptography: Fully homomorphic encryption, Bar-Ilan University, http://crypto.biu.ac.il/2nd-biu-winter-school.

[5] Gentry, C. & Halevi, S. (2010). Implementing Gentry's fully homomorphic encryption, Cryptology e-Print Archive: Report 2010/520.

[6] Halevi, S. (2013). On fully homomorphic encryption, http://www.math.uci.edu/~asilverb/Lattices/.

[7] Hussain M. & Hussain, M. (2013). A Survey of Image Steganography Techniques, International Journal of Advanced Science and Technology Vol. 54.

[8] Katzenbeisser, S. & Petitcolas, F. A. P. (2000). Information Hiding Techniques for Steganography and Digital Watermarking, Artech House, INC.

[9] Rudine, W. (1991). Functional analysis. McGraw-Hill, 1.

[10] Goluch, S. (2010). The development of homomorphic cryptography From RSA to Gentry's privacy homomorphism.

[11] Todd, M.J. & Yıldırım, E.A. (2007). On Khachiyan's algorithm for the computation of minimum-volume enclosing ellipsoids Discrete Applied Mathematics 155.

[12] Von Ahn, L. & Hopper, N. J. (2004, April). Public-key steganography. In EUROCRYPT (Vol. 3027, pp. 323-341).

[13] N. Smart & F. Vercauteren, Fully homomorphic SMID operations, http://eprint.iacr.org/2011/133, 2011.

[14] G. Swain & S.K. Lenka, 2014. Classification of image steganography techniques in spatial domain: a study. IJ.

Of Computer Science & Engineering Technology, 5(3), pp.219-232.

[15] M. Li, M.K. Kulhandjian, D.A Pados, S.N. Batalama, and M.J.Medley, 2013. Extracting spread-spectrum hidden data from digital media. IEEE transactions on on information forensics and security, 8(7), pp.1201-1210

[16] S.K. Bandyopadhya., 2011. Genetic Algorithm Based Substitution Technique of Image Steganography. Journal of Global Research in Computer Science, 1(5).

[17] , L.M. Marve, , C.G. Boncelet and C.T. Retter, 1999. Spread spectrum image steganography. IEEE Transactions on image processing, 8(8), pp.1075-1083.

[18] S. Dumitrescu, X. Wu, and Z. Wang, 2003. Detection of LSB steganography via sample pair analysis. IEEE transactions on Signal Processing, 51(7), pp.1995-2007.